\documentclass[conference]{IEEEtran}
\IEEEoverridecommandlockouts
\usepackage{cite}
\usepackage{amsmath,amssymb,amsfonts}
\usepackage{algorithmic}
\usepackage{graphicx}
\usepackage{textcomp}
\usepackage{xcolor}
\def\BibTeX{{\rm B\kern-.05em{\sc i\kern-.025em b}\kern-.08em
    T\kern-.1667em\lower.7ex\hbox{E}\kern-.125emX}}
    
\def\R{{\mathbb R}}
\def\C{{\mathbb C}}    
\def\Sp{{\mathbb S}}  
\def\Fab{{F_{a,b}}}

\usepackage{todonotes}

\begin{document}


\title{An inexact matching approach for the comparison of plane curves with general elastic metrics
\thanks{Nicolas Charon and Yashil Sukurdeep are supported by NSF grant n$^o$ 1819131. M. Bauer was partially supported by NSF-grant n$^o$ 1912037 (collaborative research in connection with NSF-grant n$^o$ 1912030)}
}

\author{\IEEEauthorblockN{Yashil Sukurdeep}
\IEEEauthorblockA{\textit{Department of Applied Mathematics} \\
\textit{Johns Hopkins University}\\
Baltimore, USA \\
yashil.sukurdeep@jhu.edu}
\and
\IEEEauthorblockN{Martin Bauer}
\IEEEauthorblockA{\textit{Department of Mathematics} \\
\textit{Florida State University}\\
Tallahassee, USA \\
bauer@math.fsu.edu}
\and
\IEEEauthorblockN{Nicolas Charon}
\IEEEauthorblockA{\textit{Department of Applied Mathematics} \\
\textit{Johns Hopkins University}\\
Baltimore, USA \\
charon@cis.jhu.edu}
}

\maketitle

\begin{abstract}
This paper introduces a new mathematical formulation and numerical approach for the computation of distances and geodesics between immersed planar curves. Our approach combines the general simplifying transform for first-order elastic metrics that was recently introduced by Kurtek and Needham, together with a relaxation of the matching constraint using parametrization-invariant fidelity metrics. The main advantages of this formulation are that it leads to a simple optimization problem for discretized curves, and that it provides a flexible approach to deal with noisy, inconsistent or corrupted data. These benefits are illustrated via a few preliminary numerical results.      
\end{abstract}


\section{Introduction}
In this article, we are interested in the computation of geodesic elastic distances between geometric curves. By geometric curves, we mean curves modulo shape preserving transformations, i.e., modulo translations, rotations and reparametrizations. Mathematically, we model the space of geometric curves as a quotient space of infinite dimensional manifolds. Although this construction involves several important technicalities, we will only rely on its basic properties for the purpose of presenting our approach. We refer interested readers to the vast literature on this topic for additional details~\cite{Bauer2014,Srivastava2016}.

Our approach combines the general simplifying transform for first-order elastic metrics, as recently introduced by Kurtek and Needham~\cite{KuNe2018}, with a relaxation of the matching constraint using parametrization-invariant fidelity metrics, resulting in an efficient implementation of the inexact elastic matching problem for planar curves. The main advantages of this formulation are that it boils down to a simple optimization problem for discretized curves, and that it provides a flexible approach to deal with noisy, inconsistent or corrupted data. These benefits are illustrated via a few preliminary numerical results. In future work, we plan to exploit the full power of this inexact matching approach as it naturally extends to elastic shape matching models that allow for partial matching constraints and topological inconsistencies. 

\section{Proposed method}
\subsection{Elastic metrics on the space of curves}
To begin the construction of elastic metrics on the space of geometric curves, we first need to model the space of parametrized curves, for which we consider the set of all smooth regular curves with values in the plane. i.e.,
\begin{equation}
\text{Imm}(M,\C):=\left\{
c\in C^{\infty}(M,\C): |c'|\neq 0
\right\}\,.
\end{equation}
The reason for representing the ambient space as the complex plane will become clear in the following section, where we introduce a key concept of this article, the $F_{a,b}$-transform. As mentioned previously, one can consider several group actions on the space of parametrized curves, notably the action of the reparametrization group $\operatorname{Diff}_+(M)$ and the actions of the groups of translations, rotations and scalings. Our analysis will primarily focus on the group of reparametrizations, which is infinite-dimensional and by far the most difficult one to handle. 

This leads us to define the space of shapes, namely curves modulo reparametrizations and translations, as follows:
\begin{equation}
\mathcal S(M,\C)=\operatorname{Imm}(M,\C)/ \{ \operatorname{Diff}(M)\times \operatorname{Tra} \}\,.
\end{equation}
To define a relevant metric on this space, we aim to follow the general setup of elastic shape analysis: 
\begin{enumerate}
    \item[i)] Define an invariant Riemannian metric on the space of regular, parametrized curves. 
    \item[ii)] By the invariance of this metric to reparametrizations and translations, it descends to a Riemannian metric on the space of shapes.
    \item[iii)] Obtain a metric between shapes from the induced geodesic distance function.
\end{enumerate}

To define a Riemannian metric on the space of parametrized curves, we note that this space carries the structure of an infinite dimensional manifold, where the tangent space is the space of all smooth functions $C^{\infty}(M,\C)$. Defining Riemannian metrics in this infinite dimensional situation can lead to unexpected phenomenons, such as vanishing geodesic distance as evidenced by Michor and Mumford~\cite{Michor2005,Bauer2012c}. We will restrict ourselves to a certain class of first order metrics, first introduced by Mio et. al~\cite{Mio2007}, which have been shown to lead to well-defined distances. For tangent vectors $h,k\in T_c\operatorname{Imm}(M,\C)$ these metrics are given by:
\begin{equation}
\begin{aligned}
    \label{eq:first_order_elastic_metrics}
    G_c^{a,b}(h,k) = \int_M a^2 \langle &D_s h, N \rangle \langle D_s k, N \rangle
    \\ &+ b^2 \langle D_s h, T \rangle \langle D_s k, T \rangle ds
    \end{aligned}
\end{equation}
where $a,b>0$ are constants, $T,N$ are the tangent and normal vector to the curve $c$, and where $ds = |c'|d\theta$, and $D_s = \frac{1}{|c'|} \frac{d}{d\theta}$ are arclength differentiation and integration respectively. The name ``elastic" refers to the fact that the first term in $G_c^{a,b}$ measures a bending energy of the curve, while the second one is associated to a stretching energy. The invariance of this family of metrics to reparametrizations follows by a straightforward application of the change of variable formula in the integral. Furthermore, these metrics are by construction insensitive to translations in the plane. The induced distance between two curves $c_0$ and $c_1$ in the quotient space $\operatorname{Imm}(M,\C) / \{\operatorname{Diff}(M) \times \operatorname{Tra} \}$ is then given by: 
\begin{equation}
    \label{eq:elastic_distance}
    d^{a,b}(c_0,c_1)^2 = \underset{\tilde{c}, \phi}{\inf} \int_0^1 G^{a,b}(\partial_t \tilde{c}, \partial_t \tilde{c})dt
\end{equation}
where the infinimum is taken over paths of immersions $t \mapsto \tilde{c}(t,\cdot) \in \operatorname{Imm}(M,\C)$ and reparametrizations $\phi \in \operatorname{Diff}(M)$, with the boundary constraints $\tilde{c}(0) = c_0$ and $\tilde{c}(1) = c_1 \circ \phi$. We point out that the formulation of the induced distance given in (\ref{eq:elastic_distance}) allows us to compute the geodesic distance $d^{a,b}(c_0,c_1)$ and the minimizing geodesic $c(t)$ as the solution of an optimal control problem. Direct numerical discretization of the functional in (\ref{eq:elastic_distance}) and its resolution is possible, but usually computationally expensive~\cite{BBHM2017,bauer2018relaxed}.

In the following section, we introduce a simplifying family of transformations, originally proposed in \cite{KuNe2018}, which reduces the above optimal control problem to a minimization problem solely on the reparametrization function $\phi$, thereby dramatically decreasing the computational complexity of solving (\ref{eq:elastic_distance}).

\subsection{The $F_{a,b}$ transform}
The beauty in the class of $G^{a,b}$ metrics lies in the fact that they allow us to derive an explicit formula for the geodesic distance on the space of parametrized, open curves. In turn, this leads to a first order approximation of the geodesic distance on the space of closed curves. This surprising fact allows one to derive extremely efficient numerical methods to solve the geodesic boundary value problem on the shape space of unparametrized curves. The first known instance of such a transformation was found for the metric with constants $a=b=1$ by Younes et. al. in~\cite{Younes1998,Michor2008a}, and for $a=1$, $b=1/2$, the celebrated SRV-framework, by Srivastava et.~al.~\cite{Jermyn2011,Srivastava2016}. These transformations were then generalized to all parameters satisfying $4a^2-b^2\geq 0$ in \cite{Bauer2014b}, and very recently, to arbitrary constants by Kurtek and Needham in \cite{KuNe2018}. In what follows, we describe the latter construction in some detail, as it is one of the two principal building blocks in our proposed approach. 

Following the presentation in \cite{KuNe2018}, we define for a regular curve $c$ the transform
\begin{equation}
\begin{aligned}
F_{a,b}: \operatorname{Imm}(M,\C) &\to C^{\infty}(M,\C)\\
c &\mapsto 2b|c'|^{1/2}\left(\frac{c'}{|c'|}\right)^{\frac{a}{2b}}\;.
\end{aligned}
\end{equation}
Here, the power in the second factor of the $F_{a,b}$-transform has to be understood in terms of complex arithmetic. The following theorem has been proven in \cite{KuNe2018} and  is the source of the importance of this transformation for our purposes:\\

\noindent
{\bf Theorem 1:} {\it The $F_{a,b}$ transform is an isometric immersion from $\left(\operatorname{Imm}(M, \C)/\operatorname{Tra}, G^{a,b}\right)$, the space of parametrized curves modulo translations, with values in the space $C^{\infty}(M,\C\}$ of smooth functions equipped with the standard $L^2$-metric. For open curves, i.e., $M=[0,1]$, the transform is a bijection onto the set of smooth curves that skip $0\in\C$.}\\

As a consequence of the above theorem, we obtain an explicit formula for geodesics and geodesic distances between open curves. Indeed, given open curves $c_0, c_1 \in \text{Imm}([0,1],\C)$, the geodesic distance between $c_0$ and $c_1$ induced by the $G^{a,b}$-metric on $\text{Imm}([0,1],\C)/\operatorname{Tra}$, also called the elastic distance, can be written as:
\begin{equation}
    \label{eq:immersion_distance}
    || \Fab(c_0) - \Fab(c_1) ||_{L^2}^2 = \int_0^1 | \Fab(c_0) - \Fab(c_1) |^2 d\theta
\end{equation}
with the associated geodesic path being given by:
\begin{equation}
\label{eq:geodesic_Fab}
    \tilde{c}(t) = F^{-1}_{a,b}((1 - t)F_{a,b}(c_0) + tF_{a,b}(c_1))\;.
\end{equation}

However, our goal is to compute distances and geodesics between $c_0$ and $c_1$ regardless of how they are parametrized, which requires the study of the distance on the space of unparametrized curves, namely, the quotient space $\operatorname{Imm}([0,1],\C) / \{\operatorname{Diff}(M) \times \operatorname{Tra} \}$. We can express the distance on this space as follows:
\begin{equation}
    \label{eq:quotient_distance}
    d^{a,b}(c_0,c_1)^2 = \underset{\phi\in \operatorname{Diff}_+([0,1])}{\inf} || \Fab(c_0) - \Fab(c_1 \circ \phi) ||_{L^2}^2\;.
\end{equation}
We point out that computing the distance $d^{a,b}(c_0,c_1)$ and its associated geodesic using (\ref{eq:quotient_distance}) reduces to optimizing over the reparametrization group $\operatorname{Diff}_+([0,1])$ only. This is a much simpler problem when compared to equation~\eqref{eq:elastic_distance}, which required solving a minimization problem over the whole path of regular curves in addition to optimizing over the reparametrization group. Consequently, we observe that~\eqref{eq:quotient_distance} provides an efficient framework to compute \textit{exact} distances and geodesics between open curves on the space of unparametrized curves $\operatorname{Imm}([0,1],\C) / \{\operatorname{Diff}([0,1]) \times \operatorname{Tra} \}$. Furthermore, for closed curves, one can use the formula for the geodesic distance between open curves as an approximation, which in turn still leads to efficient algorithms.

This framework readily adapts to piecewise linear curves, which implies that numerical solutions for~\eqref{eq:quotient_distance} can be computed efficiently by discretizing the curves $c_0, c_1$ and the objective functional in~\eqref{eq:quotient_distance}, and then optimizing over the reparametrization group $\text{Diff}_+(M)$ using a dynamic programming approach as in \cite{Mio2007}, which has a complexity of the order of $O(N^2)$, where $N$ is the number of vertices of the discretized curves.

\subsection{Varifold fidelity terms}
The second key component of our approach is the relaxation term, which we use to measure discrepancy between curves. Since the goal is to compute distances and geodesics between geometric curves by enforcing an approximate matching of one curve to another (modulo reparametrizations, translations and rotations), it is fundamental for such relaxation/discrepancy terms to be independent of the parametrization of either curve. In other words, we seek discrepancy terms that only depend on the geometric image of the curves. However, unlike the elastic metrics introduced earlier, these discrepancy terms do not need to be tied to Riemannian metrics on shape spaces, but should instead be as simple as possible to compute in practice. 

One efficient approach used for similar purposes in past works on diffeomorphic registration involves representing shapes, like curves, as objects in special spaces of measures, such as currents \cite{Glaunes2008} or varifolds \cite{Charon2013,Charon2017}. This allows one to quantify shape discrepancy by instead comparing the associated measures. Various families of distances can be considered for such purposes, including metrics derived from optimal transport, but more explicit ones can be constructed through the framework of reproducing kernel Hilbert spaces (RKHS). 

In the following paragraphs, we shall only give a very brief summary of the construction of varifold discrepancy metrics. We refer to the recent presentations of \cite{Charon2017} or \cite{Charon2020} for further details and extensions of this model.  

Given a parametrized planar curve $c \in \operatorname{Imm}(M,\C)$, we may associate to it a varifold $\mu_c$, which is specifically the measure on the product space $\C \times \Sp^1$ defined for any continuous test function $\omega: \C \times \Sp^1 \rightarrow \R$ by:
\begin{equation}
    \label{eq:var_curve}
    \left(\mu_c | \omega \right) = \int_{M} \omega\left(c(\theta),\frac{c'(\theta)}{|c'(\theta)|}\right) ds.
\end{equation}

Note that $\mu_c$ essentially corresponds to the arclength measure along the curve $c(M)$, together with its unit tangent vector $c'(\theta)/|c'(\theta)| \in \Sp^1$. We point out that $\mu_c$ does not depend on the parametrization of $c$, in the sense that for any $\phi \in \text{Diff}^+(M)$, one has $\mu_{c \circ \phi} = \mu_c$.
One can therefore compare two given curves $c_1$ and $c_2$ modulo reparametrization by comparing the varifolds $\mu_{c_1}$ and $\mu_{c_2}$. In particular, kernel metrics are well-suited for our purpose as they lead to explicit expressions of the resulting distance. Indeed, taking a positive definite kernel on $\C \times \Sp^1$ of the form $k(x,u,y,v) = \rho(|x-y|) \phi(u\cdot v)$, where $\rho$ and $\gamma$ define respectively a radial kernel on $\C$ and a zonal kernel on $\Sp^1$, we can construct a (pseudo-)metric $\|\cdot\|_{\operatorname{Var}}$ on measures of $\C \times \Sp^1$ which takes the following explicit form:
\begin{multline}
    \label{eq:var_distance}
\|\mu_c\|_{\operatorname{Var}}^2 = \iint_{M\times M} \rho(|c(\theta) - c(\theta')|)\\ \gamma\left(\frac{c'(\theta)}{|c'(\theta)|} \cdot \frac{c'(\theta')}{|c'(\theta')|} \right) ds ds'.
\end{multline}
Then, $D_{Var}(c_1,c_2)^2 := \|\mu_{c_1} - \mu_{c_2}\|_{Var}^2$, which we call the varifold fidelity metric, defines a discrepancy term between the two curves $c_1$ and $c_2$ modulo reparametrizations. The specific properties of $D_{\operatorname{Var}}$ crucially depend on the choice of kernel functions $\rho$ and $\gamma$: a more thorough discussion of this topic can be found in \cite{Charon2017} and \cite{bauer2018relaxed}. We also note that for the general class of kernels defined above, the resulting discrepancy term $D_{\operatorname{Var}}$ is equivariant to the action of translations and rotations, namely that for any $\alpha \in [0,2\pi)$ and $z \in \C$, we have $D_{\operatorname{Var}}(e^{i\alpha}c_1 + z, e^{i\alpha}c_2+z)=D_{\operatorname{Var}}(c_1,c_2)$.  

\subsection{Relaxed formulation of the geodesic problem}
\label{ssec:relaxed_formulation}
While the geodesic problem described in \eqref{eq:quotient_distance} is theoretically and numerically appealing, it relies on an exact matching of the template curve $c_0$ to the target curve $c_1$, which may be undesirable in certain practical applications. For instance, if the target curve $c_1$ is corrupted by noise, one would obtain highly inaccurate estimates of distances and geodesics by enforcing an exact matching of the template curve to the noisy version of $c_1$. Such practical concerns motivate the introduction of a relaxed formulation of the variational problem in \eqref{eq:quotient_distance}, which we describe in the next paragraphs.


Going back to the original formulation of \eqref{eq:elastic_distance}, the idea is to relax the boundary constraint $c(1) = c_1 \circ \phi$ by using the varifold fidelity metric $D_{\operatorname{Var}}$, which was introduced in the previous section. This is indeed valid, because under adequate choices of kernels, c.f. \cite{bauer2018relaxed}, one has that $D_{\operatorname{Var}}(c(1), c_1) = 0$ if and only if $c(1)$ and $c_1$ are equal up to reparametrizations. Therefore, we may rewrite the variational problem in \eqref{eq:elastic_distance} as:
\begin{equation*}
  \underset{\tilde{c}}{\inf} \int_0^1 G^{a,b}(\partial_t \tilde{c}, \partial_t \tilde{c})dt \quad \text{s.t} \quad D_{\operatorname{Var}}(\tilde{c}(1),c_1) = 0 \,.
\end{equation*}
Now, setting $c=\tilde{c}(1) \in \operatorname{Imm}(M,\C)$, it follows from Theorem 1 that the minimum of the above functional simplifies to $\|F_{a,b}(c_0) -  F_{a,b}(c)\|_{L^2}^2$. If in addition, we relax the boundary constraint with a Lagrange multiplier $\lambda > 0$, we are led to the following inexact matching problem:
\begin{equation}
    \label{eq:elastic_matching_problem}
    \underset{c \in \operatorname{Imm}(M,\C)}{\inf} || \Fab(c_0) - \Fab(c) ||_{L^2}^2 + \lambda D_{\operatorname{Var}}(c,c_1)^2 \,.
\end{equation}
 We point out that this new formulation involves optimizing over the end curve $c=\tilde{c}(1)$ only, in stark contrast with~\eqref{eq:elastic_distance}, where the minimization is over a full path of immersions as well as reparametrizations. The geodesic between $c_0$ and the approximate matched curve $c$ can be then recovered from \eqref{eq:geodesic_Fab}. However, the minimization space remains typically larger compared to the exact $F_{a,b}$ matching approach given by \eqref{eq:quotient_distance}. Yet, one important advantage of \eqref{eq:elastic_matching_problem} is the flexibility and robustness provided by the relaxation of the boundary constraint. Indeed, it allows us to adapt the weighting factor $\lambda$ to the data, which is highly desirable in some applications, as we shall illustrate in the experiments,. 
 
 \begin{figure*}
\begin{tabular}{cccc}
    \includegraphics[trim = 25mm 10mm 25mm 5mm ,clip,width=0.21\textwidth]{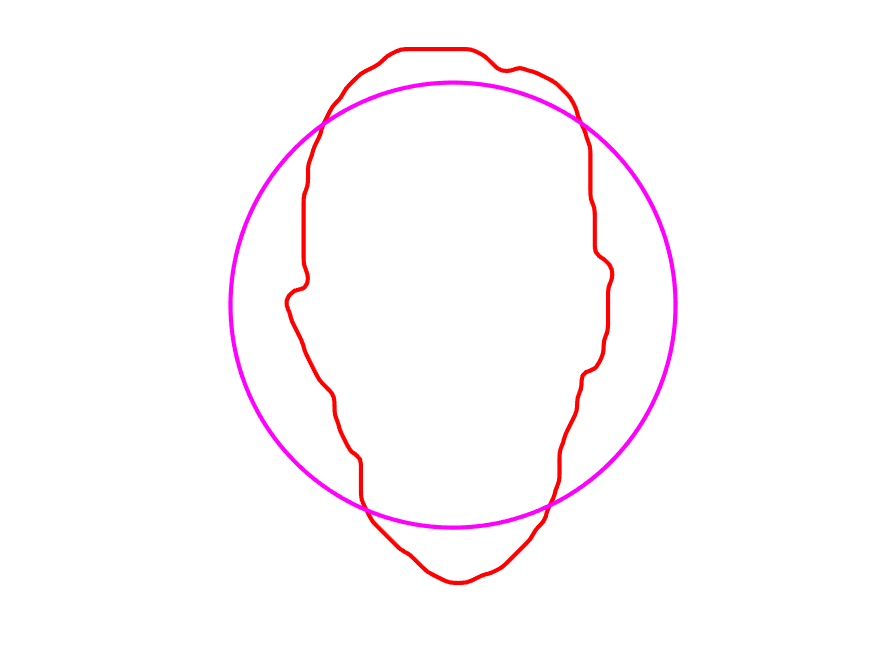} & \includegraphics[trim = 25mm 10mm 25mm 5mm ,clip,width=0.21\textwidth]{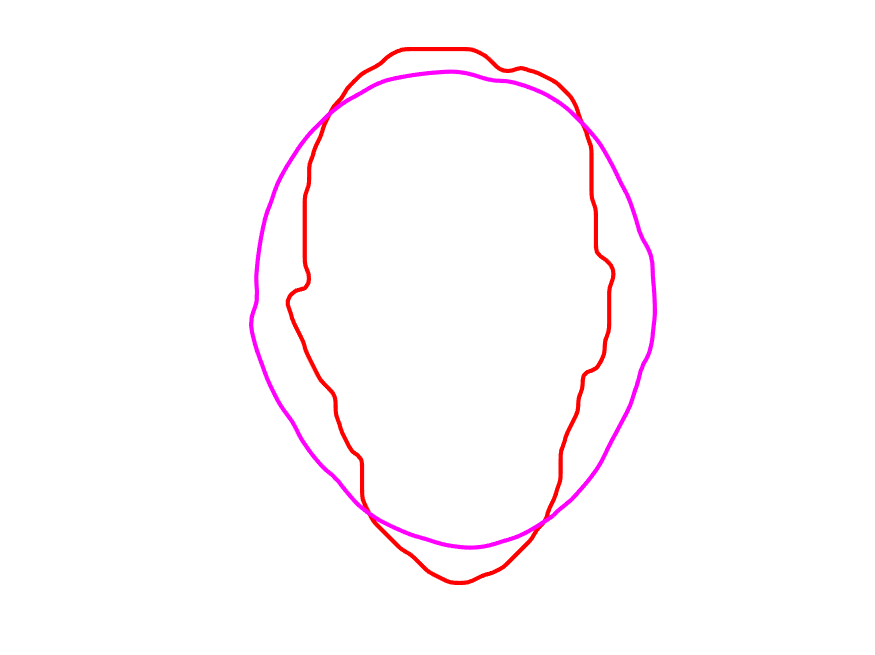} & 
    \includegraphics[trim = 25mm 10mm 25mm 5mm ,clip,width=0.21\textwidth]{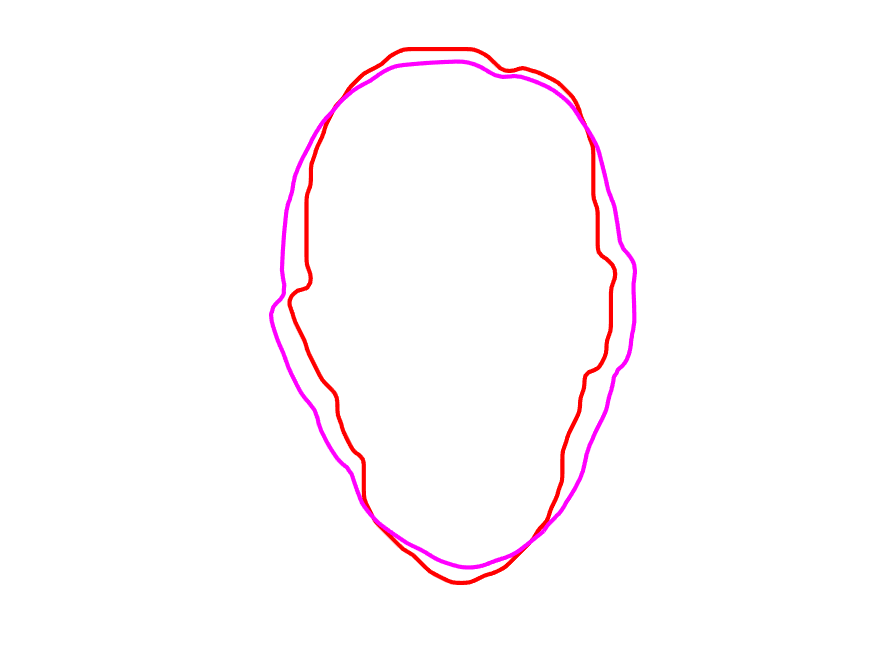} &
    \includegraphics[trim = 25mm 10mm 25mm 5mm ,clip,width=0.21\textwidth]{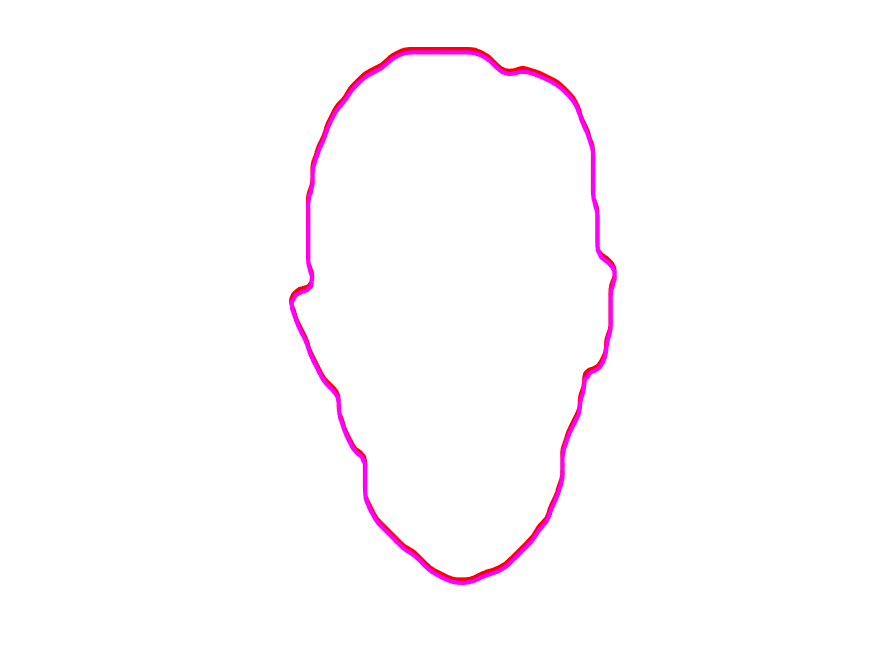} \\
    $t=0$ & $t=1/3$ & $t=2/3$ & $t=1$
\end{tabular}
    \caption{Geodesic between a circle and the red target curve obtained for $a=1$, $b=0.8$ and $\lambda =1000$.}
    \label{fig:geodesic_circle_face}
\end{figure*}

\begin{figure*}
\begin{tabular}{cccc}
    $\lambda =10$ & $\lambda=20$ & $\lambda=1000$ & Exact matching \\
    \includegraphics[trim = 25mm 10mm 25mm 5mm ,clip,width=0.21\textwidth]{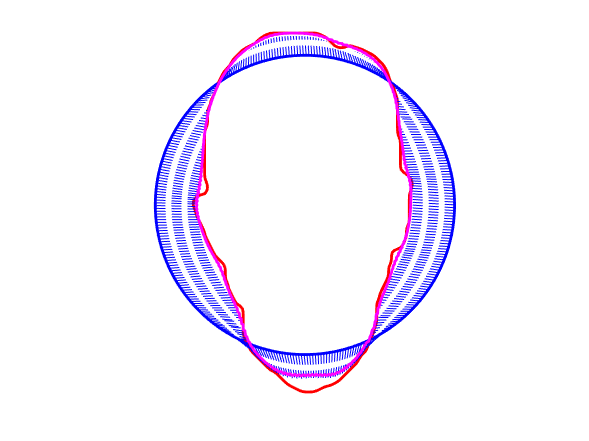} & 
    \includegraphics[trim = 25mm 10mm 25mm 5mm ,clip,width=0.21\textwidth]{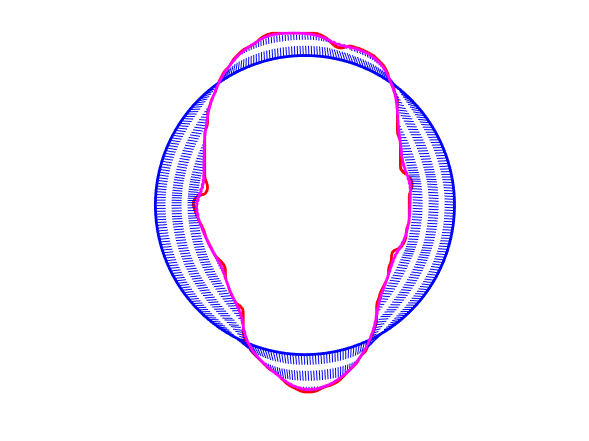} & 
    \includegraphics[trim = 25mm 10mm 25mm 5mm ,clip,width=0.2\textwidth]{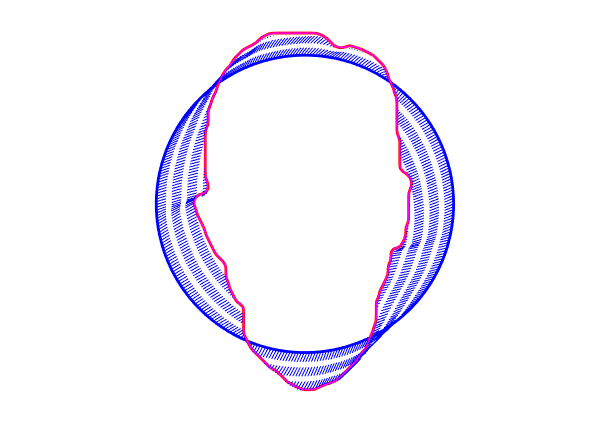} &
    \includegraphics[trim = 25mm 10mm 25mm 5mm ,clip,width=0.21\textwidth]{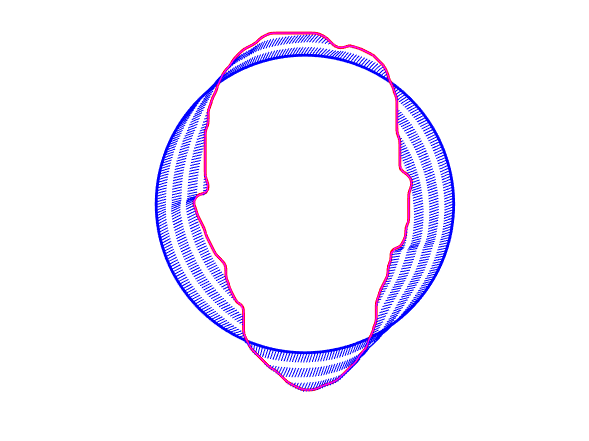} \\
    $d^{a,b} = 0.1152$ & $d^{a,b} = 0.1289$ & $d^{a,b} = 0.1412$ & $d^{a,b} = 0.1436$ 
\end{tabular}
    \caption{Effect of $\lambda$ on the geodesic path and estimated distance. The last column shows the result obtained with the exact matching algorithm of \cite{KuNe2018}.}
    \label{fig:geodesic_circle_face_lambda}
\end{figure*}
 
 Furthermore, it is fairly straightforward to specify \eqref{eq:elastic_matching_problem} in the case of piecewise linear curves; the interested reader may refer to \cite{KuNe2018} regarding the discretization of the $F_{a,b}$ term, while discretizations of varifold terms are thoroughly examined in \cite{Charon2020}. This allows us to turn~\eqref{eq:elastic_matching_problem} into a finite dimensional minimization problem over the position of the vertices of the final curve $c$. In practice, we use a limited memory BFGS algorithm to minimize the discretized functional in \eqref{eq:elastic_matching_problem}, with initializations like $c=c_0$ or $c=c_1$ depending on the application at hand.
 
 
Finally, in addition to translations and reparametrizations, it is also possible to further quotient out rotations in this framework, which is necessary in certain applications. Due to the equivariance properties of both the elastic distance and the varifold fidelity metric with respect to the action of rotations, one can modify~\eqref{eq:elastic_matching_problem} to quotient out by rotations as follows:
 \begin{equation}
    \label{eq:elastic_matching_rotations}
    \underset{c,\alpha}{\inf} ||\Fab(c_0) - \Fab(c) ||_{L^2}^2 + \lambda D_{\operatorname{Var}}(e^{i\alpha}c,c_1)^2
\end{equation}
where the minimization is now over both $c \in \operatorname{Imm}(M, \C)$ and the rotation angle $\alpha \in [0,2\pi)$. 

\section{Experimental data and results}
We now present a few results of geodesic distance computation using the relaxed framework presented in Section \ref{ssec:relaxed_formulation}. 
\vskip2ex
\noindent \textbf{A simple example.} We start with a simple example to provide a basic comparison of our method with an exact matching method, and to illustrate the effect of some of the model parameters. Fig. \ref{fig:geodesic_circle_face} shows a reconstructed geodesic evolution between two curves, obtained using our proposed approach, with elastic parameters $a=1$, $b=0.8$, and a large value of $\lambda=1000$ for the weighting parameter. This enforces a close matching to the target curve, and thus, the resulting geodesic and distance $d^{a,b}(c_0,c)=|| \Fab(c_0) - \Fab(c) ||_{L^2}$ obtained using our approach is comparable to the geodesic and distance $d^{a,b}(c_0,c_1)$ obtained with the exact approach of \cite{KuNe2018}, as shown in Fig. \ref{fig:geodesic_circle_face_lambda}. We also observe that decreasing $\lambda$ leads to a less precise matching and smaller elastic distance. \\

\begin{figure}[h!]
\begin{tabular}{cc}
    \includegraphics[trim = 25mm 10mm 25mm 5mm ,clip,width=0.2\textwidth]{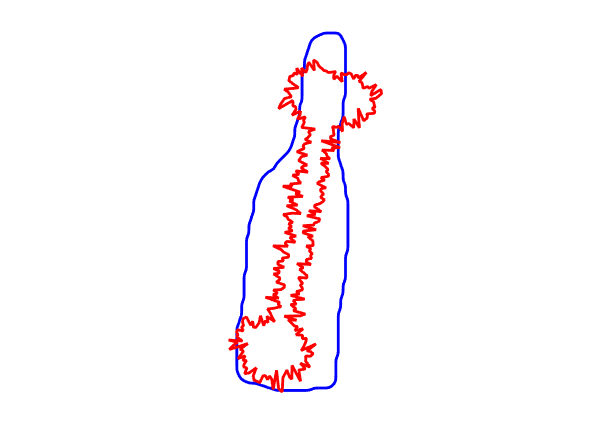} & \includegraphics[trim = 25mm 10mm 25mm 5mm ,clip,width=0.2\textwidth]{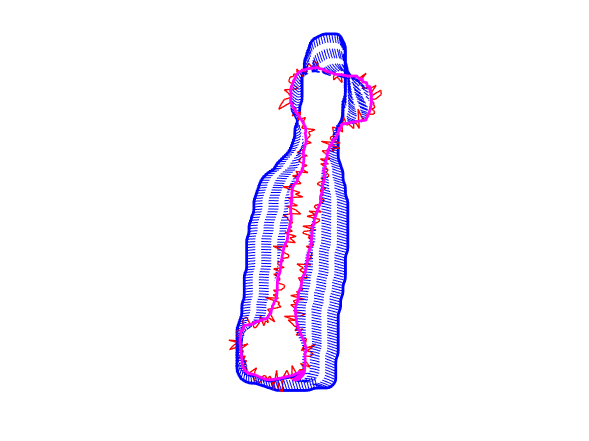}
\end{tabular}
    \caption{Left: source (blue) and noisy target curve (red). Right: estimated geodesic with our proposed approach for $a=1$, $b=0.5$ and $\lambda=40$. }
    \label{fig:matching_bottle_bone}
\end{figure}

\noindent \textbf{Noisy curves.} One possible advantage of our relaxed framework is the ability to estimate meaningful elastic distances under noise. We illustrate this in Fig. \ref{fig:matching_bottle_bone}, where the target curve is corrupted by noise. In this example, the exact elastic distance given by the algorithm of \cite{KuNe2018} equals $d^{a,b}(c_0,c_1) = 0.5358$, which is unreasonably high, mainly due to the irregularity of the target curve. Our algorithm, in contrast, estimates an end curve $c$ which is essentially an approximate and regularized version of $c_1$, leading to a distance $d^{a,b}(c_0,c) = 0.1078$, which is a more accurate estimate of the actual distance to the noise free version of the target curve. 


\begin{figure}[h!]
\begin{tabular}{cc}
    \includegraphics[width=0.24\textwidth]{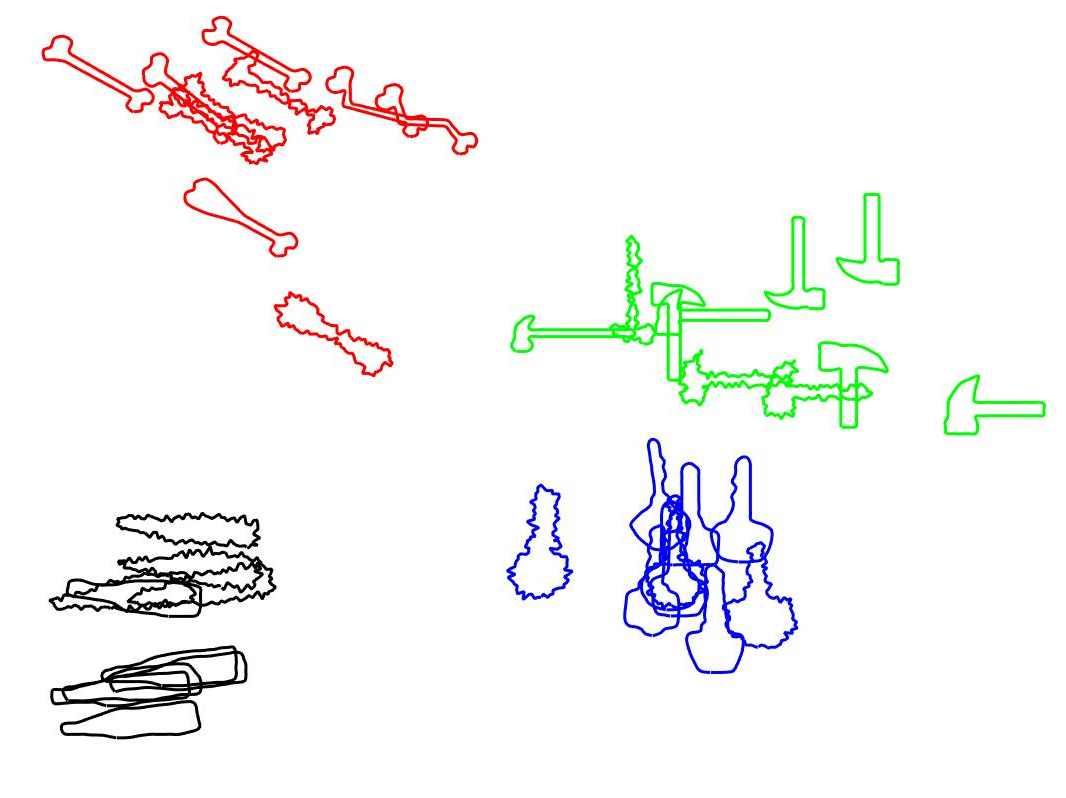} & \includegraphics[width=0.20\textwidth]{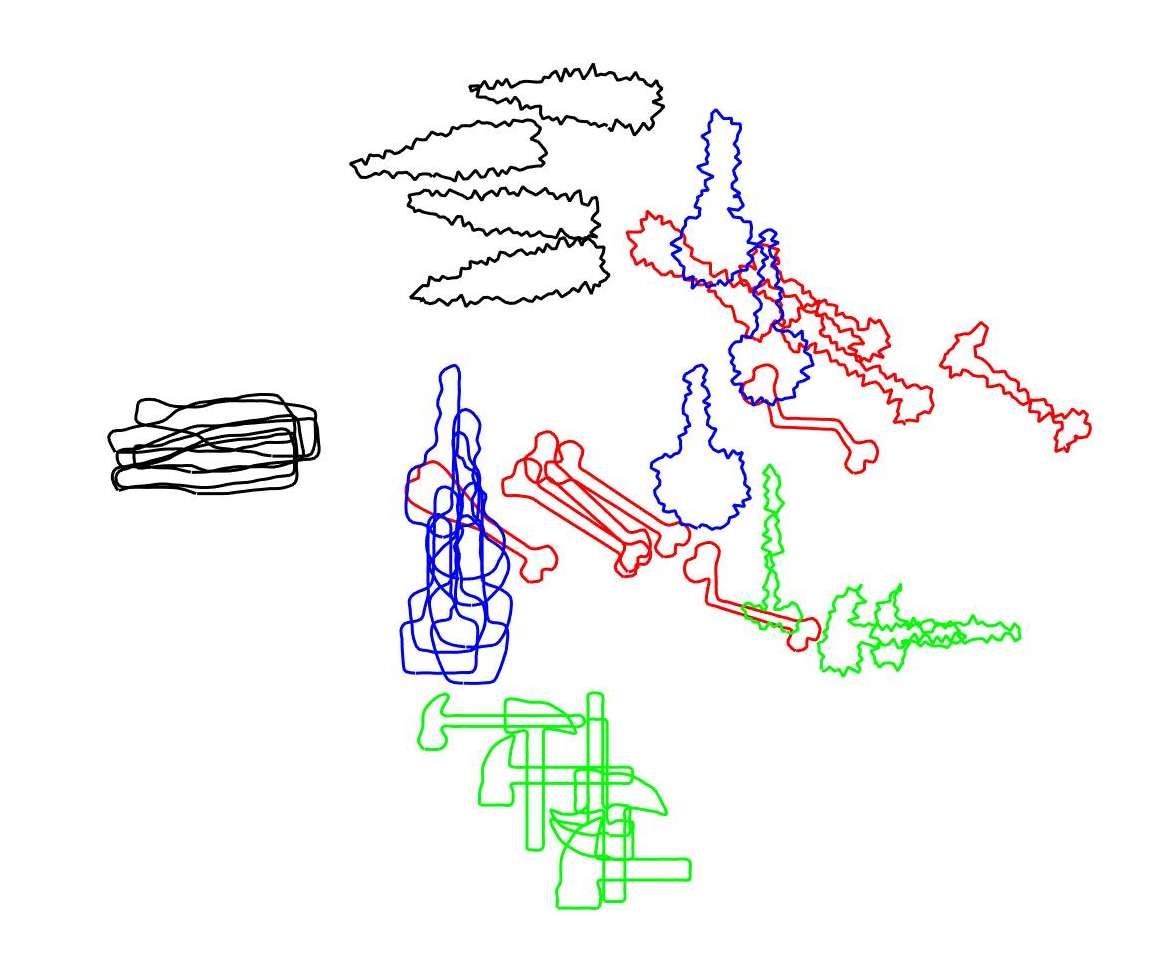}
\end{tabular}
    \caption{Multidimensional scaling plots obtained with the proposed relaxed approach (left) and the exact matching approach (right).}
    \label{fig:clustering}
\end{figure}

\noindent \textbf{Clustering comparison.} The ability to perform robust estimation of distances in the presence of noise may in turn improve statistical analysis methods based on elastic distances. We illustrate this on a simple unsupervised clustering task, performed on a set of 40 shapes selected from the Kimia database, some of which have been corrupted by noise. The chosen dataset consists of four different categories of shapes, namely bones, bottles, hammers and keys. To cluster the shapes, we compute all the pairwise (rotation-invariant) distances between them, using both the exact matching and our relaxed approach. We then apply the classical multi-dimensional scaling method to the resulting pairwise distance matrices to project the dataset onto a two-dimensional space. The resulting projections are shown in Figure \ref{fig:clustering}, where one can observe that the presence of noise has a negative impact on the quality of clusters obtained using the exact matching approach, while the relaxed approach leads to much more consistent clusters.      

\begin{figure}
\begin{tabular}{cc}
    \includegraphics[width=0.22\textwidth]{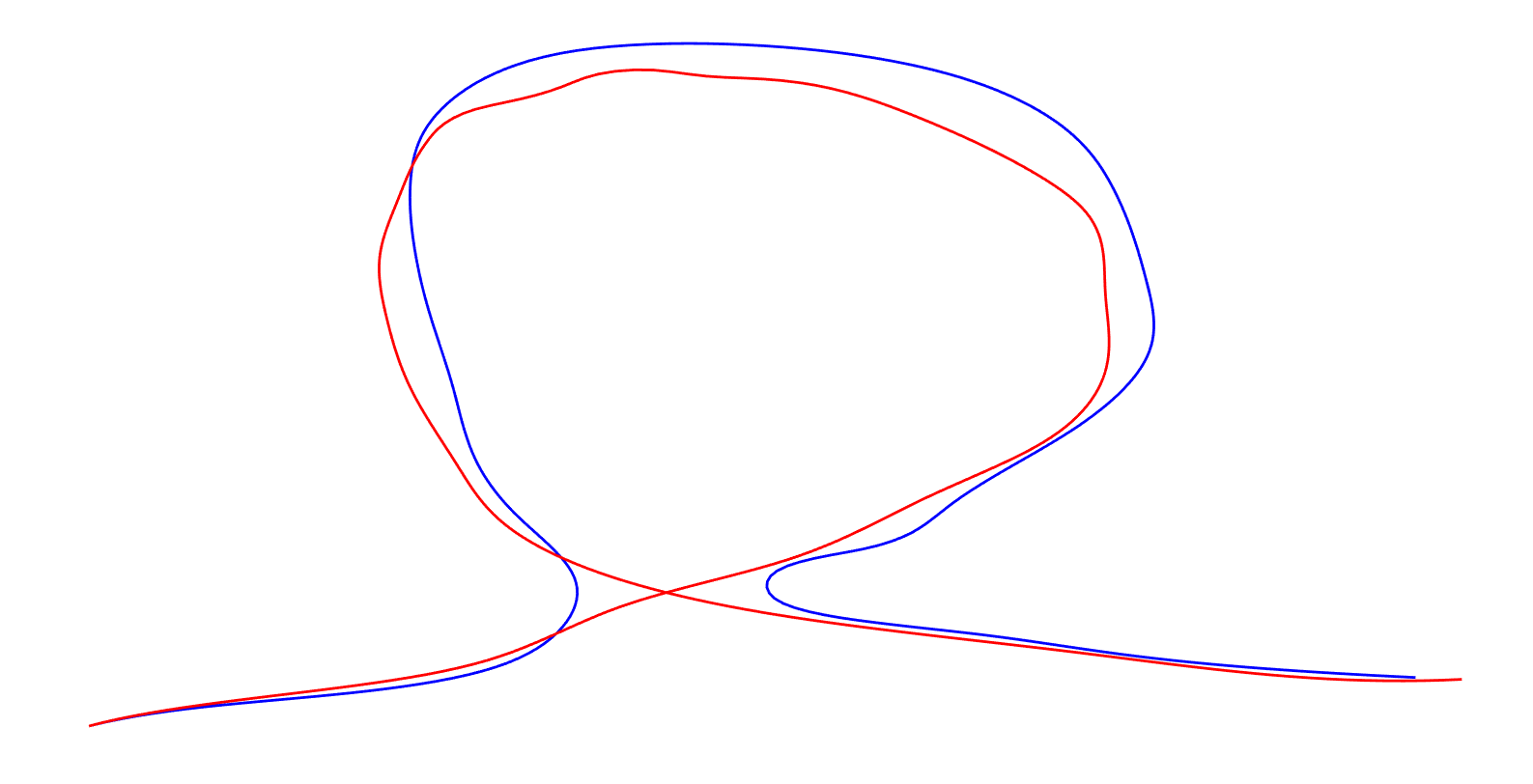} & \includegraphics[width=0.22\textwidth]{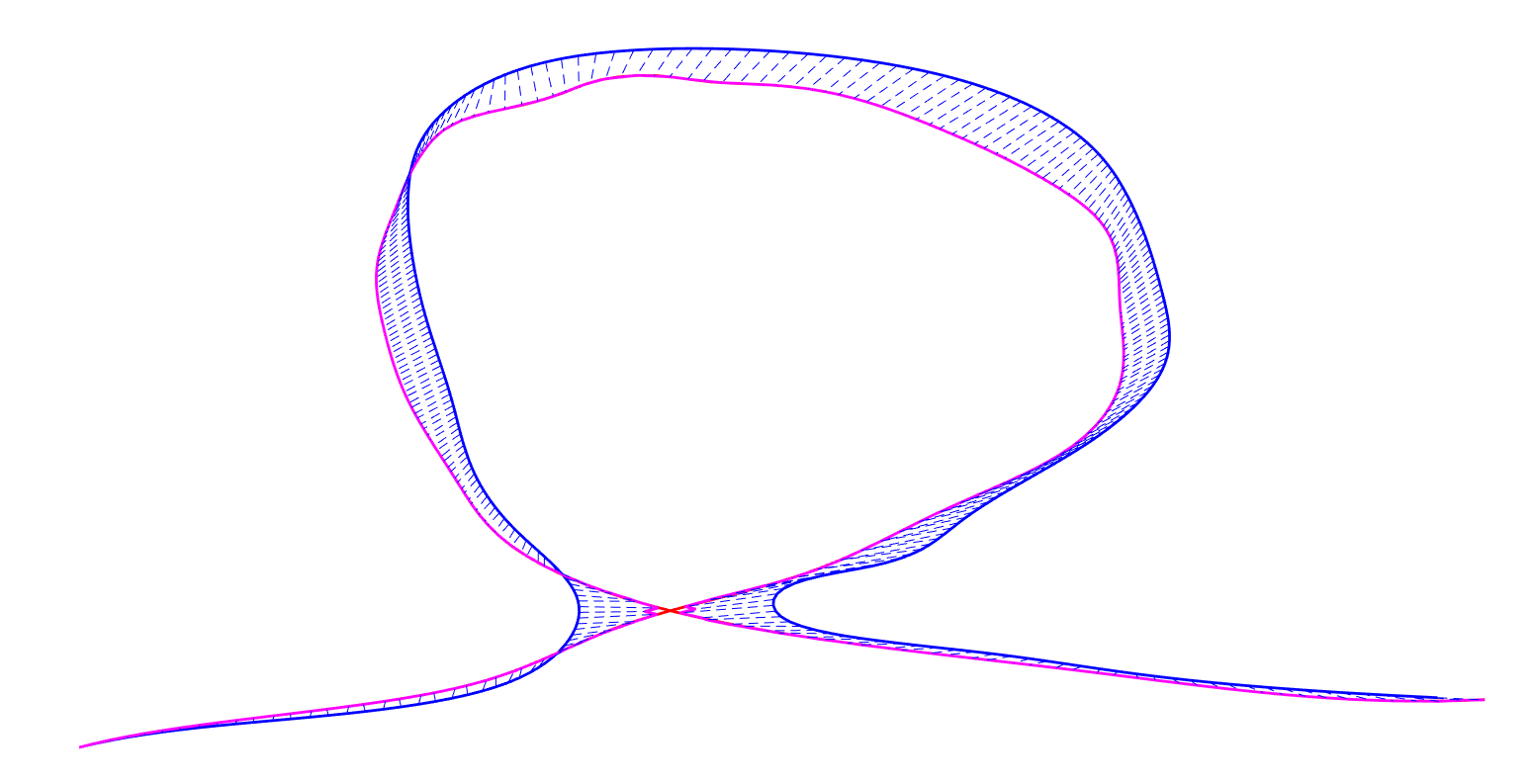}
\end{tabular}
    \caption{Left: example of two curves that have similar geometric images, but have different topologies. Right: estimated matching by our algorithm.}
    \label{fig:matching_topology}
\end{figure}

\noindent \textbf{Topological noise.} One last interesting feature of our formulation of the elastic matching problem is that it provides the possibility to compare curves which exhibit small topological variations. This is enabled by the varifold discrepancy terms, which are quite flexible in dealing with curves of different topologies and/or orientations. Fig. \ref{fig:matching_topology} shows an example borrowed from \cite{KuNe2018}, where the two curves, despite being very close in shape, are topologically not equivalent. This may occur for instance due to inconsistencies or imprecisions in the segmentation process. This small topological difference induces a large elastic distance in the exact matching setting, where one finds that $d^{a,b}(c_0,c_1)=3.9$. However, by relaxing the constraint, in this case, using an orientation-invariant instance of varifold metric, we recover a rather natural matching which leads to a significantly smaller elastic distance $d^{a,b}(c_0,c)=0.027$.    


\section{Conclusion and further extensions}
We have proposed an inexact reformulation of the elastic matching problem for planar curves, which takes advantage of both the simplification provided by the $F_{a,b}$ transform, and the versatility of varifold-based discrepancy terms. Our approach provides a robust way to deal with geodesic distance computations in the presence of noise and perturbations. Moreover, we expect that this approach could be extended to the more challenging situation of immersed surfaces: this has so far only been touched upon for some very specific choice of metric in \cite{bauer2019inexact}. Another promising avenue, which is the subject of ongoing work by the authors, is to leverage the flexibility of the varifold representation for modelling and estimating weight functions defined on the shapes. This can allow us to incorporate partial data matching constraints, which have seldom been considered in this elastic metric framework, and could possibly enable the joint modelling of elastic and topological variations.


\bibliographystyle{abbrv}
\bibliography{references}

\end{document}